\pdfoutput=1 
\documentclass[acus]{JAC2003}

\usepackage{graphicx}
\usepackage{booktabs}

\newcommand{\pa}{\partial}

\newcommand{\ket}{\rangle }
\newcommand{\bra}{\langle }

\newcommand{\dket}{\ket\ket}
\newcommand{\dbra}{\bra\bra}
\newcommand{\up}{\uparrow}
\newcommand{\dw}{\downarrow}

\newcommand{\Vect}[1]{\mbox{\boldmath$#1$}}

\begin{document}
\title{Strong field physics in condensed matter\thanks{Work supported by Grant-in-Aid for 
Scientific Research on Priority Area
``New Frontier of Materials Science Opened by Molecular Degrees of Freedom".
}}

\author{T. Oka\thanks{oka@cms.phys.s.u-tokyo.ac.jp}, University of Tokyo, Japan
}

\maketitle

\begin{abstract}
There are deep similarities between 
non-linear QFT studied in high-energy and 
non-equilibrium physics in condensed matter.
Ideas such as the Schwinger mechanism
and the Volkov state are deeply related to 
non-linear transport and photovoltaic Hall effect in condensed matter. 
Here, we give a review on these relations.

\end{abstract}
\section{Introduction}

\begin{figure}[bth]
\centering 
\includegraphics[width=7.cm]{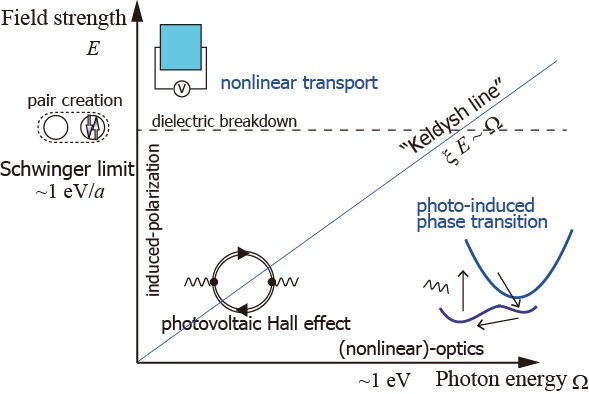}
\caption{
Several phenomena in condensed matter physics in 
strong electric fields plotted in
the $(E,\Omega)$-space.
}
\label{fig:overview}
\end{figure}

In strong field physics, researchers are interested in 
the change of the ``quantum vacuum" due to strong external fields. 
A typical example is the decay of the QED vacuum in strong 
electric fields due to the 
Schwinger mechanism \cite{Schwinger1951}. 
When a strong enough electric field is applied 
to the vacuum, pair creation of electrons and positrons 
takes place and the insulation break down. 
The threshold for this phenomenon is known as 
Schwinger's critical field and is given by $E_{\rm th}
=m_{\rm e}^2/{\rm e}=1.3\times 10^{16}\mbox{V/cm}$.
Since the critical field is extremely strong,
direct experimental verification 
is still a challenge.  
On the other hand, in the condensed matter community, 
there is an increasing interest in 
non-equilibrium phase transitions 
and non-linear transport in strongly correlated electron systems
(Fig.~\ref{fig:overview})\cite{Oka_LZreview,
Oka2003,Oka2005a,OkaSchwinger10,PhysRevLett.105.146404,OkaPHE}. 
In the experiments, one also applies 
strong electric fields
and the original insulating phase is destroyed. 
However, the threshold for dielectric breakdown 
is orders of magnitude smaller than the Schwinger mechanism 
in QED since the excitation gap is far smaller. 
This makes condensed matter systems to be an idealistic 
playground to test and develop theoretical ideas in non-linear QFT.
Non-linear physics has been studied rather independently
in the two fields, high energy and condensed matter, 
during the past few decades, and several parallel ideas were developed. 
The aim of this article is to explain
some of the correspondences (Table~\ref{correspondence}).

\begin{table*}[tb]
   \centering
   \caption{Related ideas in strong field physics}
   \begin{tabular}{cc}
       \toprule
       \textbf{High Energy} & \textbf{Condensed Matter}  \\ 
       \midrule
Schwinger mechanism in QED&Landau-Zener tunneling in band insulators\\
Heisenberg-Euler effective Lagrangian&
Non-adiabatic geometric phase, Loschmidt Echo\\
Vacuum polarization&
Extended Berry's phase theory of 
polarization\\
Pair creation in interacting systems (e.g. QCD)&
Many-body Schwinger-Landau-Zener 
mechanism \\
&in strongly correlated system\\
Dirac particles in circularly polarized light
&Photovoltaic Hall effect\\
Furry picture& Floquet picture \\
\bottomrule
   \end{tabular}
   \label{correspondence}
\end{table*}

\section{Pair creation in strong electric fields}
\subsection{Hesenberg-Euler's effective action and 
the non-linear extension of
the Berry's phase approach to polarization \cite{Oka2005a}:}
We study lattice electrons 
in homogeneous electric fields. 
In the time-dependent gauge, this can be 
realized by adding a time dependent phase 
to the hopping term in the lattice Hamiltonian.
For example, for a one-dimensional model,
a typical Hamiltonian is given by
\begin{eqnarray}
H(\Phi)&=&-\sum_{i=1}^L\sum_\sigma(e^{i\Phi} c_{i+1\sigma}^\dagger c_{i\sigma}
+e^{-i\Phi}c^\dagger_{i\sigma}c_{i+1\sigma})\\
&&+U\sum_{i}n_{i\up}n_{i\dw}+\sum_{i}V_in_{i}.\nonumber
\end{eqnarray}
We impose periodic boundary condition 
and the phase $\Phi$ is proportional to the 
magnetic flux through the ring ($L$ number of sites).
The time derivative of the magnetic flux is related to the 
applied electric field through $F(t)=eaE(t)=d\Phi(t)/dt$,
where $e$ is the charge quantum and $a$ the lattice constant. 
$U$ represents on-site Coulomb repulsion 
and $V_i$ the local potential. The hopping term
is set to unity. 
The Hubbard model ($U>0,\;V_i=0$) at half-filling 
is in the Mott insulating phase for 
positive $U$ in one dimension. 

Here, we study what happens to the an insulator when we 
apply strong electric fields. 
We denote the eigenstates
of the Hamiltonian $H(\Phi)$ by
$|\psi_n(\Phi)\ket,\;n=0,1,\ldots$
and study the time evolution 
starting from the groundstate $|\psi_0(\Phi)\ket$.
The groundstate-to-groundstate amplitude defined by
\begin{eqnarray}
\Xi(t)\equiv\bra \psi_0(\Phi(t))|e^{-i\int_0^tH(\Phi(s))ds}|\psi_0(0)\ket 
e^{i\int_0^tE_0(\Phi(s))ds}
\end{eqnarray}
is of central importance. 
In the long time limit, 
an asymptotic behavior ($d$ is dimension)
$
\Xi(t)\sim e^{itL^d\mathcal{L}}
$
is expected to take place where
$\mathcal{L}$ is the condensed matter
version of the {\it Heisenberg-Euler effective Lagrangian}.
The imaginary part describes quantum tunneling
where
$
\Gamma(F)/L^d \equiv 2\mbox{Im}\;\mathcal{L}(F)
$
gives the speed of the exponential decay 
of the vacuum (groundstate).
This quantity is proportional to the decay rate of the Loschmidt Echo $L(t)=|\Xi(t)|^2$. 
The real part $\mbox{Re}\mathcal{L}$ is
written in terms of a non-adiabatic phase called the
Aharonov-Anandan phase (which we denote $\gamma$) that the
wave function acquires during the 
time-evolution. For band insulators ($U=0$)
in dc-electric fields, the effective
Lagrangian becomes \cite{Oka2005a}
\begin{eqnarray}
\mbox{Re}\;\mathcal{L}(F) &=&-F
\int_{\rm BZ} \frac{d\Vect{k}}{(2\pi)^{d}}\frac{\gamma
(\Vect{k})}{2\pi},\\
\mbox{Im}\;\mathcal{L}(F) &=&-F\int_{\rm BZ} 
\frac{d\Vect{k}}{(2\pi)^{d}}\frac{1}{4\pi}\ln \left[ 1-p(\Vect{k})\right],
\label{2Schwinger2}
\end{eqnarray}
where the momentum integral is over the 
Brillouin Zone (BZ).
There is an interesting parallel theory developed in the 
condensed matter community. 
This is known as the Berry's phase 
theory of polarization\cite{Resta1992,KingSmith1993,Resta1998,Resta1999,Nakamura2002}, where 
the ground-state expectation value of the twist operator 
$e^{-i\frac{2\pi}{L}\hat{X}}$, which shifts the phase of electron wave 
functions 
on site $j$ by $-\frac{2\pi}{L}j$, plays a crucial role. 
It was revealed that the real part of a quantity 
\begin{equation}
w=\frac{-i}{2\pi}\ln\bra 0|e^{-i\frac{2\pi}{L}\hat{X}}|0\ket
\end{equation}
gives the electric polarization
$P_{\rm el}=-\mbox{Re}w$ \cite{Resta1998}
while itfs imaginary part gives a criterion for metal-insulator
transition, i.e.,
$D=4\pi\mbox{Im}w$ is finite in insulators 
and divergent in metals \cite{Resta1999}.
The present effective Lagrangian can be regarded as 
a non-adiabatic (finite electric field) extension of $w$.
To give a more accurate argument, 
we rewrite the effective Lagrangian in the time-independent gauge
\begin{equation}
\mathcal{L}(F)\sim \frac{-i\hbar}{\tau L}\ln\left(
\bra 0|e^{-\frac{i}{\hbar}\tau(H+F\hat{X})}
|0\ket e^{\frac{i}{\hbar}\tau E_0}\right)
\end{equation}
for $d=1$. 
Let us set $\tau=h/LF$ and consider the small $F$ limit.  
For insulators we can replace $H$ with the
groundstate energy $E_0$ to have 
$\mathcal{L}(F)\sim wF$ 
in the linear-response regime. Thus the real part of 
Heisenberg-Euler's expression\cite{Heisenberg1936} for 
the non-linear polarization 
$P(F)=-\pa \mathcal{L}(F)/\pa F$ 
naturally reduces to the Berry's phase 
formula $P_{\rm el}$ in the small field limit $F\to 0$.
Its imaginary part 
gives the criterion for photo-induced 
metal-insulator transition, originally proposed
for the zero field case.

\subsection{Many-body Schwinger-Landau-Zener mechanism in strongly correlated insulators:}

\begin{figure}[b]
\centering 
\includegraphics[width=7.cm]{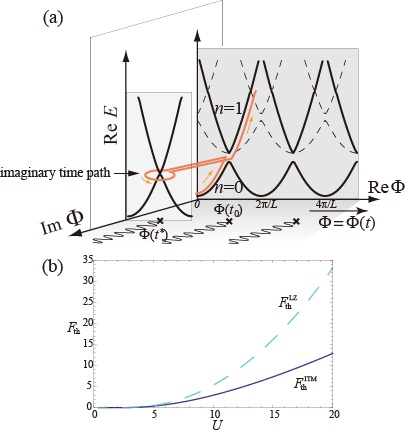}
\caption{(a)
Many-body energy levels against the complex AB flux $\Phi$ 
for a finite, half-filled 1D Hubbard model 
($L=10,\;N_\up=N_\dw=5$, $U=0.5$).  
Only charge excitations are plotted. 
Quantum tunneling occurs between the groundstate (labeled as $n=0$) 
and a low-lying excited state ($n=1$)
as the flux $\Phi(t)=Ft$ increases 
on the real axis, while 
the tunneling is absent for the 
states plotted as dashed lines.  
The wavy lines starting from the singular points 
($\times$) at $\Phi(t^*)$ 
represent the branch cuts for different Riemann surfaces, 
along which the solutions $n=0$ and $n=1$ are connected. 
In the DDP approach, the tunneling factor is 
calculated from the dynamical phase associated
with adiabatic time evolution (DDP path) 
that encircles a gap-closing  point at
$\Phi(t^*)$ on the complex $\Phi$ plane. 
(b) Threshold electric field obtained by
the imaginary time method (solid) and the 
naive Landau-Zener formula (dashed).}
\label{fig:slz}
\end{figure}
Next, let us consider dielectric breakdown 
in a strongly correlated system. In
the one-dimensional Mott insulator where the groundstate 
is a state with one electron per site,
the relevant charge excitations 
are doublons, i.e,. doubly occupied sites,
and holes, i.e., sites with no electron. 
Pairs of doublons and holes play 
a similar role as the pair of electrons and 
positrons in the Schwinger mechanism. 
Indeed, it has been shown that dielectric breakdown in Mott insulators 
takes place due to pair production of charge excitations 
through quantum tunneling, which is
called the many-body Schwinger-Landau-Zener mechanism 
\cite{Oka2003,Oka2005a,OkaSchwinger10}. 
The tunneling probability for the 
lowest excited state in the 
one-dimensional Hubbard model was recently calculated 
by combining the imaginary time method
with the exact Bethe ansatz solution \cite{OkaSchwinger10}. 

The imaginary time method 
has been widely used in quantum physics, e.g.,
in quantum chemistry \cite{Dykhne1962,DavisPechukas1976}
and in high energy \cite{Brezin1970,PopovSovJNP1974,PopovJETP1972}. 
In high energy, the tunneling rate for 
Dirac particles in ac-electric fields was calculated
by analytically extending the time-evolution to 
complex time processes. 
This method can be readily applied to the one-dimensional Hubbard model
since the wave functions, including the excited states, 
are analytically known. Using Bethe ansatz, these 
states can be described as a linear combination 
of plane waves with the momentum (rapidity) 
determined by the Lieb-Wu equation \cite{Hubbardbook}.
In the infinite size limit, the Lieb-Wu equation 
can be solved even in the presence of a
complex phase, i.e., $\Phi\to \Phi+i\Psi$ \cite{Fukui,OkaSchwinger10}.
In the half-filled Hubbard model, 
the lowest energy excited state 
is given by the string solution
\cite{PhysRevB.9.2150,Ovchinnikov70,Takahashi72,Woynarovich1982}.
We denote the energy difference between this
state and the groundstate by $\Delta=E_{1}-E_{0}$.
For dc-electric fields, quantum tunneling to the lowest 
energy state dominates and one can express the decay rate by
\begin{equation}
\Gamma/L\sim -\frac{\alpha F}{2\pi}\ln[1-\exp(-\pi F_{\rm th}/F)],
\label{eq:gamma}
\end{equation}
where $F_{\rm th}$ is the threshold field for this process
and $\alpha$ a numerical factor. 
In the imaginary time method, an adiabatic evolution to a 
gap closing point in the complex $\Phi+i\Psi$ space is performed
and the tunneling probability depends on the dynamical phase. 
The gap closing in complex $\Phi$ has been studied in ref.~\cite{Fukui}, while
in ref.~\cite{Stafford1993} a related problem was 
studied in order to calculate the localization length in a Mott insulator. 
The Bethe ansatz solution for the Hubbard model with finite $\Psi$ 
is represented by a rapidity distribution with end points $\pm a+ib$. 
The imaginary part $b$ is related to $\Psi$ by a condition 
that the charge quantum number should remain real.  
The threshold for the tunneling is given by\cite{OkaSchwinger10}
\begin{eqnarray}
&&F_{\rm th}^{\rm ITM}
 = \frac{2}{\pi}\int_0^{b_{\rm cr}}\Delta\frac{d\Psi}{db}
db\nonumber\\
&=& 
\frac{2}{\pi}\int_0^{\sinh^{-1}u}
4\left[u-\cosh b+\int_{-\infty}^\infty
d\omega\frac{e^{\omega\sinh b}J_1(\omega)}
{\omega (1+e^{2u|\omega|})}
\right]
\nonumber\\
&&\times\left[
1-2\cosh b
\int_0^\infty d\omega\frac{J_0(\omega)\cosh(\omega \sinh b)}{1+
e^{2 u\omega}}\right]db.
\label{eq:Fthbethe}
\end{eqnarray}
Its $U$-dependence is plotted e in Fig.~\ref{fig:slz}~(b) 
(solid line). This indicates a 
collective nature of the breakdown 
(i.e., the threshold much smaller than a naive 
$U$). In other words, the tunneling 
takes place not between neighboring sites, 
but over an extended region
due to a {\it leakage of the many-body wave function},
where the size is roughly given by
the localization length \cite{Stafford1993}. 
One can compare this result with the naive estimate for the 
tunneling threshold obtained by the Landau-Zener 
formula \cite{Oka2003,Oka2005a}
\begin{eqnarray}
F_{\rm th}^{\rm LZ}=\frac{(\Delta/2)^2}{v},
\label{eq:Flz}
\end{eqnarray}
where $\Delta$ is the charge gap (Mott gap) \cite{Lieb:1968AM}, 
and $v$ is the slope of the adiabatic levels 
($v\sim 2$ when $U$ is small and the system size is small). 
From Fig.~\ref{fig:slz}~(b), one notice that the 
result of the imaginary time method 
and the Landau-Zener formula coincides in the 
small $U$ limit. However, when the 
interaction becomes stronger, large deviation takes place,
with different asymptotic behaviors
 $F_{\rm th}^{\rm ITM}\propto U-\mbox{const.}$
and $F_{\rm th}^{\rm LZ}\propto U^2$. 
Dielectric breakdown in the Hubbard
model has been  studied numerically
and threshold behaviors in the decay rate \cite{Oka2005a}
as well as in the 
current\cite{Heidrich-Meisner10,PhysRevLett.105.146404} are observed. 
Further research to explore the 
strong $U$ limit is yet to be done.

In ac-electric fields, the tunneling behavior can still be realized. 
The Keldysh line in Fig.~\ref{fig:overview} 
corresponds to the line where the 
adiabaticity parameter \cite{Keldysh65} $\tilde{\gamma}=\Omega/F\xi$
crosses unity ($\xi$ is the localization length in the present case). 
For $\tilde{\gamma}\gg 1$, the tunneling process
is the dominant cause of carrier production,
while when $\tilde{\gamma}\ll 1$, multiphoton 
absorption wins.
In the next section, another interesting effect of the 
ac-electric field is discussed where light can be used to 
control the topology of the band structure.

\section{Photovoltaic Hall effect}
\begin{figure}[tbh]
\centering 
\includegraphics[width=7.cm]{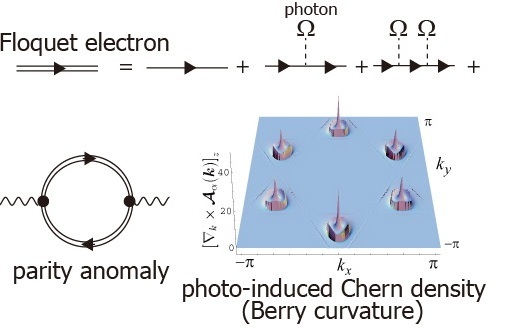}
\caption{Fig.5 (upper) Greenfs function in the Floquet picture. (lower left) Diagram for conductivity. (lower right) Photo-induced Berry curvature (Chern density) of graphene in circularly polarized light. }
\label{fig:phe}
\end{figure}
Dirac electrons are now becoming one of the central 
topics in condensed matter after its experimental 
realization in graphene\cite{Novoselov10222004}. 
Electrons in strong background electric fields
is an important problem and can be experimentally 
studied by optical and transport methods. 
In non-linear QED, this problem has been studied by Volkov  
for ac-fields where he found that electron wave functions 
acquire a non-trivial phase\cite{Volkov35}. 
However, in condensed matter, the 
electrons subject to strong lasers behave 
differently from the Volkov state of 
relativistic Dirac particles. 
This is because
the velocity, or the ``speed of light", of 
Dirac electrons in condensed matter systems 
are much slower compared to the actual speed of light 
and the ac-field breaks the 
``Lorentz symmetry" of the electron. 
This leads to a dynamical gap opening in the 
pseudo-energy diagram \cite{syzranov:045407,OkaPHE}. 
Especially, in a circularly polarized light, 
a gap opens at the Dirac point \cite{OkaPHE}.
This has an important physical consequence since a 
gap of a 2+1 dimensional Dirac electron is 
related to parity anomaly and is detectable through transport measurements,
i.e., the Hall effect. 
In 2+1 dimension, the Hall conductivity
can be written as a momentum integral of the 
Berry curvature ($\sim$ Chern density) over the Brillouin zone.
This is known as the TKNN formula\cite{tknn}, and is know 
extended to ac-driven transport via the Floquet picture
($\sim$ Furry picture) \cite{OkaPHE}
\begin{eqnarray}
\sigma_{xy}(\Vect{A}_{\rm ac})=e^2\int \frac{d\Vect{k}}{(2\pi)^d}\sum_\alpha
f_\alpha(\Vect{k})\left[\nabla_{\Vect{k}}\times\Vect{\mathcal{A}}_\alpha(\Vect{k})\right]_z.
\label{eq:tknn}
\end{eqnarray}
Here, $\Vect{\mathcal{A}}_\alpha(\Vect{k}) \equiv 
-i\dbra\Phi_\alpha(\Vect{k})|\nabla_{\Vect{k}}|\Phi_\alpha(\Vect{k})\dket$
is the photo-induced artificial gauge field.
In the Floquet picture, the Green's function
incorporates the effect of 
photon absorption and emission (Fig.~\ref{fig:phe}~(a)), 
and Hall conductivity is given by the bubble diagram 
in the non-interacting case, which is nothing but the 
parity anomaly diagram. The photo-induced Berry curvature
shown in Fig.~\ref{fig:phe}~(c) 
acts as an artificial magnetic field and 
becomes finite when the circularly polarized light is introduced. 

\begin{figure}[tb]
\centering 
\includegraphics[width=7.cm]{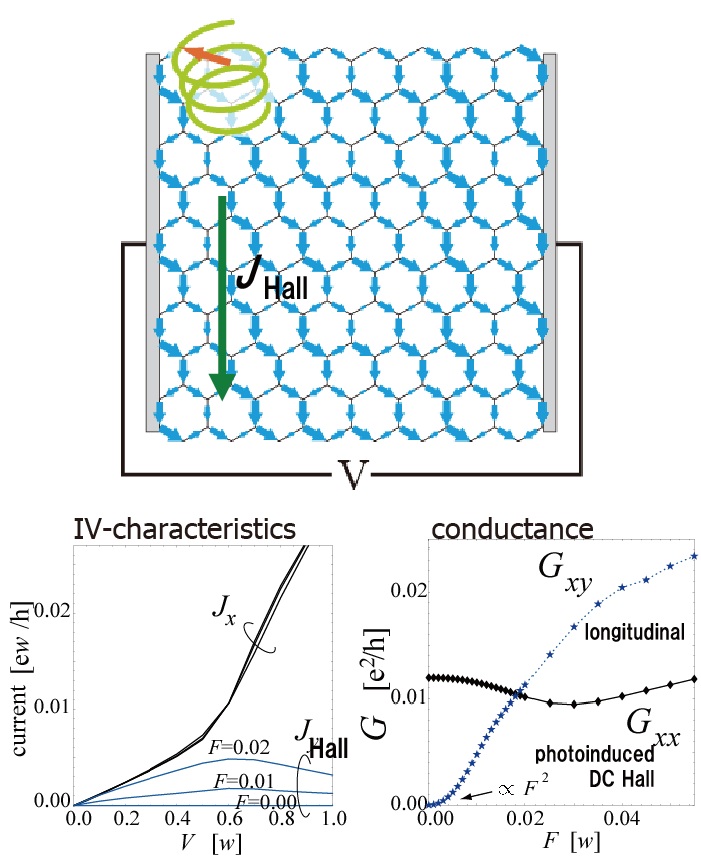}
\caption{(upper) DC current. (lower left) IV-characteristics. (lower right) Conductance.}
\label{fig:phe2}
\end{figure}

The current in the presence of  
circularly polarized light in a graphene
ribbon attached to two electrodes is plotted in 
Fig.~\ref{fig:phe2}.
The calculation has been done by combining the 
Keldysh green's function method with the 
Floquet picture.  
The Hall current, which is originally absent,
increases as the strength of light becomes stronger.
The numerical result supports our understanding of the
photovoltaic Hall effect obtain by the extended 
TKNN formula (eqn.(\ref{eq:tknn})).

We would like to acknowledge Naoto Tsuji, Martin Eckstein
and Philipp Werner for enlighting discussions. 
It is a pleasure to thank Gerald Dunne 
for illuminating discussions during PIF2010.


\end{document}